\documentclass[aps,prl,superscriptaddress,reprint,floatfix]{revtex4-2}
\usepackage{float}
\usepackage{amsmath}
\usepackage{graphicx}
\usepackage{longtable}
\usepackage{colortbl}
\usepackage[english]{babel}
\usepackage{array}
\usepackage{multirow}
\usepackage{tabularx}
\usepackage{booktabs}
\usepackage{bm}
\usepackage{svg} 
\usepackage{natbib}
\usepackage[colorlinks,bookmarks=false,citecolor=blue,linkcolor=red,urlcolor=blue]{hyperref}
\graphicspath{{graphs/}}
\newcommand{\FT}[1]{{\color{black}#1}}
\usepackage[commentmarkup=todo,deletedmarkup=sout,highlightmarkup=uuline,authormarkup=superscript,addedmarkup=colored]{changes}

\begin{document}

\title{Optical absorption spectroscopy probes water wire and its ordering in a hydrogen-bond network}
\author{Fujie Tang} 
\affiliation{Department of Physics, Temple University, Philadelphia, PA 19122, USA}
\affiliation{Pen-Tung Sah Institute of Micro-Nano Science and Technology, Xiamen University, Tan Kah Kee Innovation Laboratory (IKKEM), Xiamen, 361005, China}

\author{Diana Y. Qiu}
\thanks{$^\ast$Corresponding author. Email: diana.qiu@yale.edu}
\affiliation{Department of Mechanical Engineering and Materials Science, Yale University, New Haven, CT 06520, USA}

\author{Xifan Wu}
\thanks{$^\ast$Corresponding author. Email: xifanwu@temple.edu}
\affiliation{Department of Physics, Temple University, Philadelphia, PA 19122, USA}
\date{\today}

\begin{abstract}
 Water wires, quasi-one-dimensional chains composed of hydrogen-bonded (H-bonded) water molecules, play a fundamental role in numerous chemical, physical, and physiological processes. Yet direct experimental detection of water wires has been elusive so far. Based on advanced $ab$ $initio$ many-body theory that includes electron-hole interactions, we report that optical absorption spectroscopy can serve as a sensitive probe of water wires and their ordering. In both liquid and solid water, the main peak of the spectrum is discovered to be a charge transfer exciton. In water, the charge transfer exciton is strongly coupled to the H-bonding environment where the exciton is excited between H-bonded water molecules with a large spectral intensity. In regular ice, the spectral weight of the charge transfer exciton is enhanced by a collective excitation occurring on proton-ordered water wires, whose spectral intensity scales with the ordering length of water wire. The spectral intensity and excitonic interaction strength reaches its maximum in ice XI, where the long-range ordering length yields the most pronounced spectral signal. Our findings suggest that water wires, which widely exist in important physiological and biological systems and other phases of ice, can be directly probed by this approach.

\end{abstract}
		
\maketitle

\section{Introduction}

\par In liquid water and crystalline ice, water molecules construct an extended hydrogen-bond (H-bond) network in the condensed phase. Besides its tetrahedral structural motif, these H-bonds can form persistent, extended networks that act as a sort of highway for charge, energy and information. These networks, known as ``water wires'', play an essential role across numerous physical, chemical and physiological processes~\cite{Xingcai1998,Dellago2003,Freier2011,Hassanali2011,Paulino2020,Li2021,Lasave2021a,Kratochvil2023}. In liquid environments, water wires under biological confinement are believed to conduct both information and nutrition in a living cell~\cite{Gerhard2007,Freier2011,Li2021,Paulino2020,Kratochvil2023}. While bulk water lacks a persistent water wire, it has been repeatedly proposed that a transient water wire facilitates proton transfer in water — a process that underlies acid-base chemistry~\cite{Halle1983,Hassanali2011,Charles2013,Bekçioğlu2015,Agmon2016,Chen2018,Walker2021,Tang2021}. In crystalline ice, the emergence of proton ordering on water wires was found to prelude the phase transition from ordinary ice to ferroelectric ice~\cite{Xingcai1998,Lasave2021a}.

\par Although playing a crucial role in facilitating dynamical processes and phase transitions, the evidence for the existence of water wires remains largely circumstantial. These structures are presumed necessary to facilitate transport processes observed experimentally in aqueous environments, yet their precise molecular structure has not been directly observed to date. This is because characterization demands that the experiment detects not only the existence of H-bonds, but also the intricate and dynamical network of correlations between H-bonds while chains are being formed. In conventional scattering experiments, only positional information can be extracted~\cite{Soper2013,Skinner2013}. Optical spectroscopy is, in principle, sensitive to both positional and angular distributions~\cite{schmidt2005optical,parson2007modern,mak2012optical}, but so far, previous works have mainly relied on X-ray absorption spectroscopy to probe the structure of the H-bond network~\cite{Wernet2003,Tse2008,Fransson2016,Smith2017,Tang2022,Tang2023}. This core-level spectroscopy generates an intramolecular exciton that is too strongly localized on the individual excited water molecules to convey detailed information about the H-bond network~\cite{Fransson2016,Smith2017,Tang2022}. Its spectrum carries little information on how the H-bonds are interconnected, and it is, thus, insensitive to the existence of water wires. Extending the characterization to optical spectroscopy, which excites the more delocalized valence electrons, has potential to offer greater access to the H-bond network.

\par Despite the potential of optical absorption as a probe of delocalized states across H-bonds, the correlation of spectral features with the underlying H-bond network is immensely complex, requiring quantitatively accurate descriptions of delocalized correlated electron-hole excitations in a theoretical environment that accurately captures both the H-bonds and the many-electron interactions present in the condensed phase. In this article, we develop such a theoretical platform and report that optical absorption spectroscopy, a widely used experimental technique, can serve as a sensitive probe of the water wire. We justify our approach by a novel assignment of the experimentally observed spectral features to the structural signature of the H-bond network. Our theoretical analyses are based on $ab$ $initio$ calculations within a many-body framework that takes into account dynamical many-electron screening effects, the single-particle self energy, and electron-hole interactions. The absorption of UV light creates charge transfer excitons across H-bonds that give rise to a distinct spectral feature around $\sim$8 eV~\cite{Hahn2005,Garbuio2006,Hermann2008,Hermann2011,Vinson2012,Garbuio2015,Ziaei2016,Nguyen2019}. In bulk liquid water, the dynamical breaking and reformation of H-bonds on a timescale of picoseconds~\cite{Emilio2006,Kumar2007,Perakis2016} broadens the spectral feature, while in regular ice, the spectral feature is much more pronounced than that in water due to the emergence of water wires with intact H-bonds. On a proton ordered water wire, the aligned electric dipoles along the ordering direction facilitate a collective excitation of charge transfer excitons leading to a dramatic enhancement of the optical transition strength. As the ordering length of the water wire increases, the excitonic effects due to the collective excitation of charge transfer excitons is enhanced as well, reaching a maximum oscillator strength and exciton binding energy in ice XI, where the water wires exhibit long-range order~\cite{Xingcai1998,Lasave2021a}.

\section{Methods}
\par Our theoretical calculations use the $ab$ $initio$ GW-BSE method within many-body perturbation theory as implemented in the BerkeleyGW software package~\cite{Hybertsen1986,Rohlfing2000,Deslippe2012}. The molecular structure is modeled by using machine-learning deep potential molecular dynamics simulations, where the model is trained at the level of the hybrid meta-GGA (SCAN0) exchange-correlation (XC) functional~\cite{zhang2021jpcb}. The quasiparticles wavefunctions are computed by hybrid density functional theory (DFT) at the level of the PBE0 XC functional~\cite{Perdew1996,Perdew1996jcp,Adamo1999}. The use of hybrid DFT functional, for both molecular and electronic structures, mitigates the errors due to self-interaction and missing derivative discontinuity~\cite{Faber2013,Caruso2014,Kaplan2015}, i.e. common drawbacks in conventional DFT approaches which lead to an artificially stronger H-bonding strength and overestimated electron-hole attractions in water and ice. 
\par The configurations of liquid water are extracted from a path-integral deep potential molecular dynamics (PI-DPMD) simulation trajectory. The deep potential model was trained on the DFT data obtained with the hybrid strongly constrained and appropriately normed (SCAN0) functional~\cite{zhang2021jpcb}. The cubic cells contain 32 water molecules while the size of supercell was adjusted to have the same density in experiment. PI-DPMD simulation was performed in the $NVT$ ensemble at T = 300 K using periodic boundary conditions with the cell sizes fixed at 9.708 $\rm \AA$. The Feynman paths were represented by 8-bead ring polymers coupled to a color noise generalized Langevin equation thermostat (i.e. PIGLET)~\cite{Ceriotti2009,Ceriotti2012}. The total length of PI-DPMD simulation used well equilibrated $\sim500$ ps long trajectories.  After we obtained the MD trajectory, we used a score function (see Sect. I in the Supplemental Materials) to pick out two representative snapshots for the following GW-BSE calculation. For the configurations used in the ice Ih calculation, we first generated 8 initial geometries of ice Ih following the \textit{ice rules}~\cite{Bernal1933,Rahman1972} with different random seeds, in which the protons are in a disorder structure. The supercell contains 64 water molecules with a size of $a = 8.9873$ $\rm \AA$, $b = 15.5665$ $ \rm \AA$, and $c = 14.6763$ $\rm \AA$. For the ice XI, the number of water molecules and cell size are set to the same as the setting used in the ice Ih. We optimized the structure of Ice XI at 0 K at the SCAN0 level of theory before used for the GW-BSE calculation. 

\par The GW-BSE calculations were done with the BerkeleyGW package~\cite{Deslippe2012}. We first constructed the wavefunctions as the starting point for our GW-BSE calculation using DFT at the level of the hybrid generalized gradient approximations (GGA) of Perdew, Burke and Ernzerhof (PBE0)~\cite{Perdew1996,Perdew1996jcp,Adamo1999} with 25$\%$ of Hartree–Fock exchange energy, as implemented in Quantum ESPRESSO package~\cite{Giannozzi2017}. We chose PBE0 to reduce the overestimated charge transfer effects of induced by the ground state theory~\cite{cohen2008i,cohen2012c}. The multiple-projector norm-conserving pseudopotentials that match the potentials for oxygen and hydrogen were generated by using the ONCVPSP package~\cite{hamann2013}. The plane-wave kinetic energy cutoff was set to 85 Ry. The BSE calculation was done with 128 valence band states and 192 conduction band states for liquid water, 192 valence band states and 240 conduction band states for ice, which are enough to cover the energy range in which we are interested. For all the BSE calculations, we solved the electron-hole excitations within the GW-BSE approach in the Tamm-Dancoff approximation (TDA)~\cite{Rohlfing2000}. Further details of the calculations can be found in \FT{the Section II of} the Supplemental Material. \FT{We also calculate the real part of the dielectric function as well as the absorption coefficient spectra of liquid water and ice, reported in the Section VII of the Supplemental Material.}

\begin{figure}[hbtp]
	\setlength{\abovecaptionskip}{0.cm}
	\centering
	\includegraphics[width=3.5in]{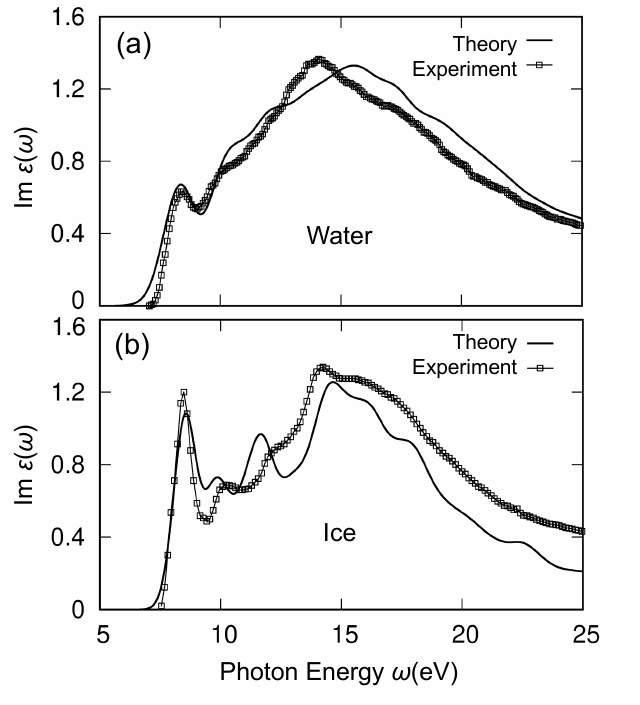}
	\caption{\label{fig:figure1} The experimental and theoretical optical absorption spectra of (a) liquid water (a) and (b) ice. The theoretical optical spectra of liquid water was generated by using the atomic configuration (32 H$_2$O) from a PI-DPMD simulation at 300K, while the ice spectra was generated by using the atomic configuration (64 H$_2$O) from a DPMD simulation for ice Ih at 80K. The theoretical spectra are aligned with experimental data with the position of first peak with a blue shift of $\sim$0.6 eV (liquid water) and $\sim$0.8 eV (ice), respectively. The experimental data are taken from Refs.~\cite{Hayashi2015} and ~\cite{Koichi1983} for water and ice, respectively. }
	\vspace{-1.5em}
	
\end{figure}

\section{Results and Discussion}
\subsection{Theoretical optical spectra of liquid water and ice}
\par We present our calculated optical absorption spectra for liquid water at 300 K and ice at 80 K in~\ref{fig:figure1} (a) and (b), respectively. The calculations are performed using the $ab$ $initio$ GW plus Bethe Salpeter equation (GW-BSE) method~\cite{Hybertsen1986,Rohlfing2000,Deslippe2012}. Details of the calculation may be found in \FT{Section I and II of the} Supplemental Material. For comparison, the recent experimental measurements from Refs.~\cite{Koichi1983,Hayashi2015} are also shown in the same figure. The theoretical water and ice spectra are blue-shifted by $\sim$0.6 eV and $\sim$0.8 eV, respectively, and both broadened by 0.4 eV with a Gaussian function to match the energy and width of the first peak in experiment. Clearly, an accurate agreement can be seen between experiments and theoretical predictions. In both water and ice, the absorption onset is dominated by a major absorption peak centered at $\sim$8 eV followed by a more broadly distributed spectral feature at $\sim$15 eV.  Despite the similarities, ice is distinct from water in its spectrum. Compared to that in water, the spectrum of ice shows a much stronger absorption peak at $\sim$8 eV together with more pronounced spectral features for the incident photon energy between $\sim$8 eV and $\sim$15 eV.

\par
\begin{figure}[t!]
	\setlength{\abovecaptionskip}{0.cm}
	\centering
	\includegraphics[width=3.5in]{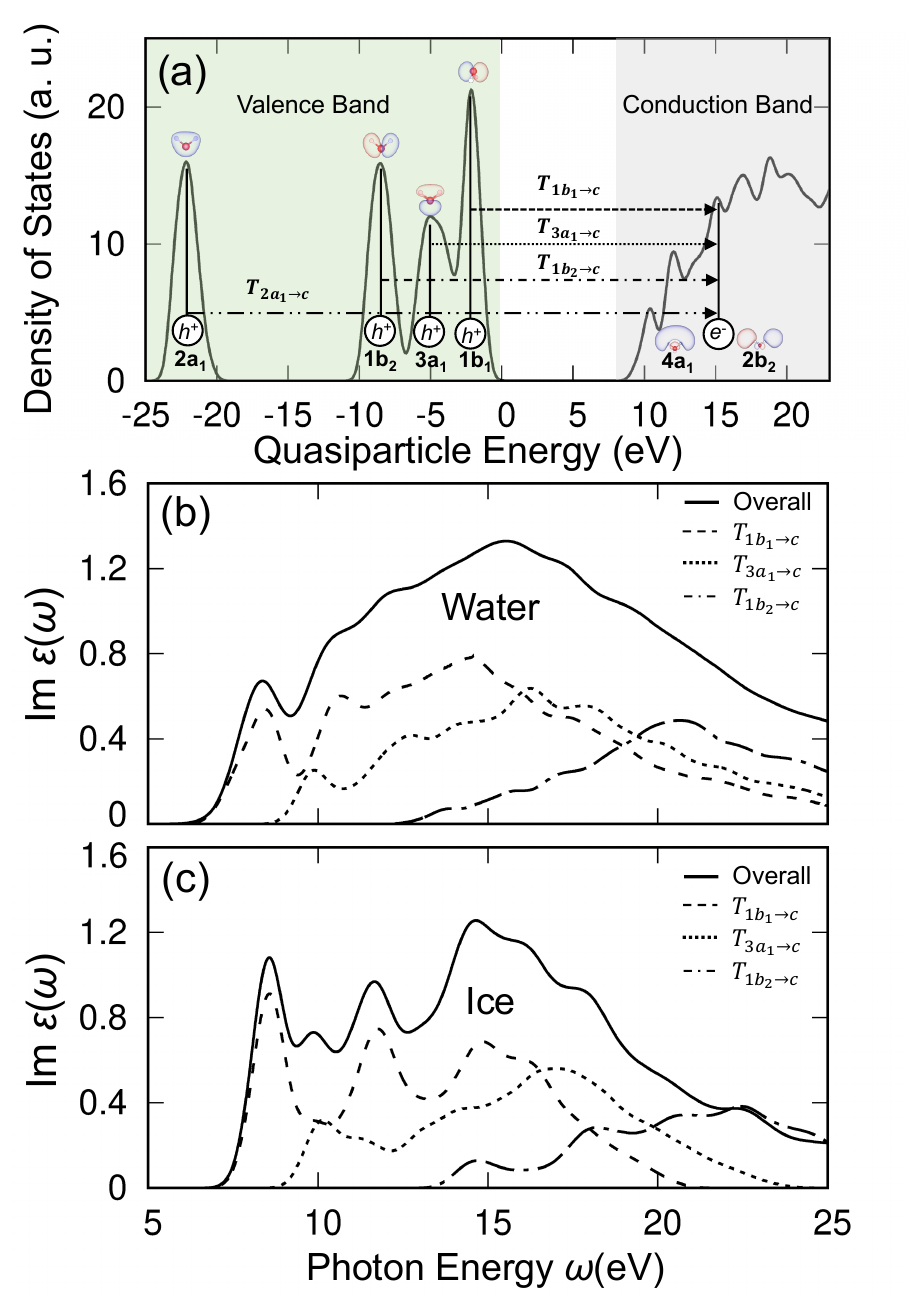}
	\caption{\label{fig:figure2} (a) Schematic of band-to-band transitions contributing to the optical spectra: ${\cal{T}}_{1b_{1}\rightarrow c}$, between valence band states with $1b_1$ characteristic and conduction band states $c$; ${\cal{T}}_{3a_{1}\rightarrow c}$, between valence band states with $3a_1$ characteristic and conduction band states $c$; ${\cal{T}}_{1b_{2}\rightarrow c}$, between valence band states with $1b_2$ characteristic and conduction band states $c$; ${\cal{T}}_{2a_{1}\rightarrow c}$, between valence band states with $2a_1$ characteristic and conduction band states $c$. The black vertical lines mark the peak of the quasi-particle density of states. The theoretical optical spectra of liquid water (b) and ice (c) from the GW-BSE approach with all valence band states contribution (solid line), and with ${\cal{T}}_{1b_{1}\rightarrow c}$ transition only (dashed line), ${\cal{T}}_{3a_{1}\rightarrow c}$ transition only (dotted line), ${\cal{T}}_{1b_{2}\rightarrow c}$ transition only (dash-dotted line). Note that the contribution from the transition ${\cal{T}}_{2a_{1}\rightarrow c}$ is zero.}
	\vspace{-1.0em}
\end{figure}

 In the condensed phase, intermolecular interactions broaden molecular orbital energy levels into bands of finite width. Optical absorption spectroscopy probes dipole-allowed transitions between these bands as shown in Fig. {\ref{fig:figure2}} (a). To understand the nature of the spectral features, we break down the absorption spectrum by contributions from specific band to band transitions. In water, the highest occupied bands are valence band states with $1b_1$, $3a_1$, $1b_2$, and $2a_1$ character, respectively, and the lowest unoccupied bands have characters $4a_1$ and $2b_2$, respectively. The schematic of the orbital character of water molecules is shown in  Fig. {\ref{fig:figure2}}(a). \FT{In the full calculation, all electron-hole pairs contribute to the spectra. We decompose the spectra into contributions from specific transitions by solving the BSE on a subspace with only selected orbitals, we find that the full spectrum is well approximated by the solution on a subspace containing only the $p$ orbital electrons near the Fermi level (i.e. the $1b_1$, $3a_1$, $1b_2$ bands), see the details in the Section IV of the Supplemental Material.}

\par In Fig. {\ref{fig:figure2}}(a), we label the transitions between the valence and conduction bands that contribute to the optical absorption. Then, in Fig. {\ref{fig:figure2}}(b), we show how each transition contributes to the overall optical absorption by restricting the calculation of the optical absorption spectrum to specific valence and conduction bands. From this, we see that in both liquid water and ice the first absorption peak around $\sim$8 eV comes from transitions from valence states of $1b_1$ (or oxygen $p$) character. The broad feature at $\sim$15 eV comes from the broadening of several distinct exciton peaks arising from transitions from $1b_1$ and $3a_1$ orbitals, because of the dispersion of the oxygen $p$ band in Fig. {\ref{fig:figure2}} (a). The low-lying oxygen $2s$ valence band of $2a_1$ character does not contribute to the optical absorption spectrum in the current energy range.

\subsection{Charge transfer exciton strongly couples to H-bond}

\par
\begin{figure*}[ht]
	\setlength{\abovecaptionskip}{0.cm}
	\centering
	\includegraphics[width=6.75in]{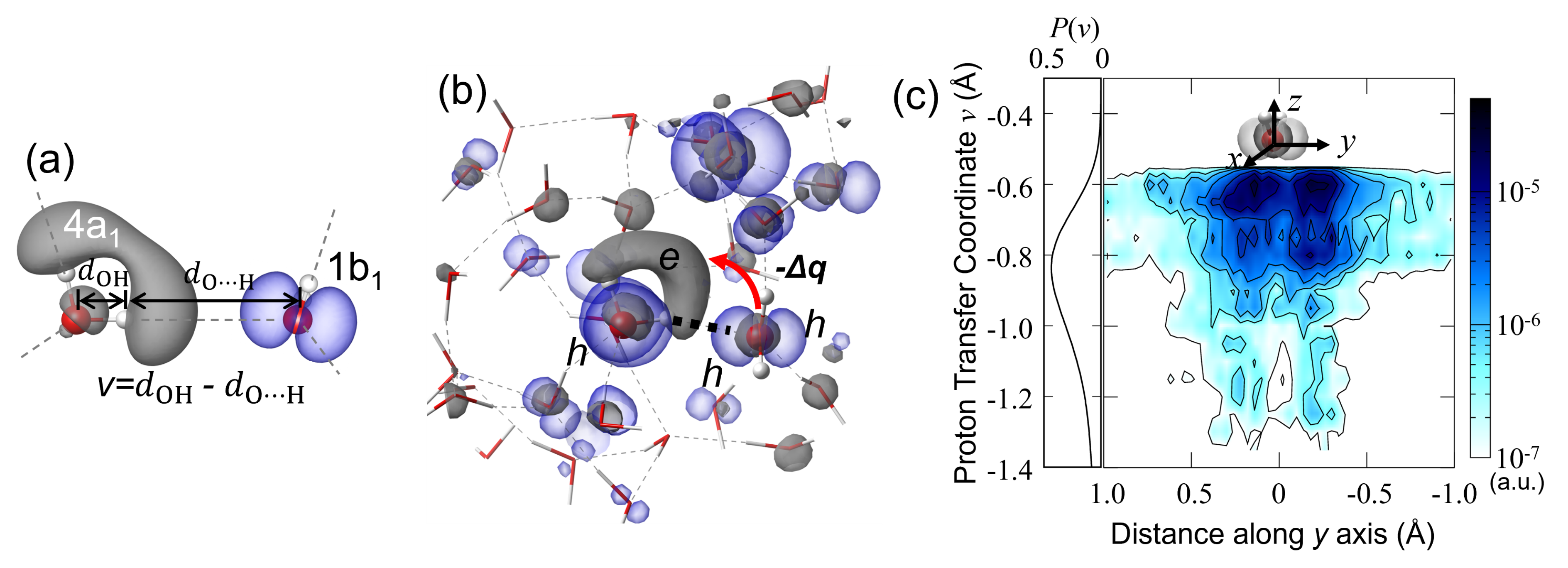}
	\caption{\label{fig:figure3} 		
		(a) The schematic of the H-bond and  proton transfer coordinate (${\nu}=d_{{\rm OH}}-d_{\rm {O\cdots H}}$).	(b) Exciton corresponding to the first peak of the optical spectrum of the liquid water in real space. The real-space plot is the averaged electron and hole density of the exciton. The blue color denotes the hole density, while the grey color denotes the electron density. (c) The two-dimensional contour plot of the overlap between hole density and electron density ($\rho_{h}\times\rho_{e}$) with respect to the distance along the y axis (perpendicular to the plane of the water molecule) and the proton transfer coordinate. The proton transfer coordinate distribution is shown for reference. The $x$-$z$ plane is parallel to the water molecule. The zero point of the $y$ axis is set at the position of oxygen atom.
	}
	\vspace{-1.0em}
\end{figure*}

  \FT{The first absorption peak at $\sim$8 eV, as shown in Fig.~\ref{fig:figure3}(a), exhibits charge-transfer characteristics, as noted in previous studies~\cite{Hahn2005,Garbuio2006,Hermann2008}. Therefore, it can be classified as a bound exciton of the charge-transfer type.} In the condensed phase, water molecules construct a tetrahedral structure via H-bonding. The H-bonding can be described by a Coulombic attraction between the electronegative oxygen on an acceptor molecule and the electropositive proton on a donor molecule, as illustrated in Fig. {\ref{fig:figure3}} (a). During the process of optical absorption, electron-hole pair excitation also occurs between the donor and acceptor molecules as shown in the Fig. {\ref{fig:figure3}} (a). An electron of $4a_1$ characteristic is excited in the lowest conduction band, which is an antibonding orbital centered on the protons of a water molecule, and at the same time, a hole of $1b_1$ character is excited in the highest valence band, which consists of lone pair electrons located in close vicinity of the oxygen atom. Not surprisingly, the resulting exciton strongly couples to the H-bonding, which applies an electrostatic field pointing from the donor to the acceptor. In the photo-absorption process, the electron in the lone pair orbitals on the H-bond acceptor molecule is excited onto the antibonding orbital on the H-bond donor molecule. Driven by the Coulombic attraction, the process of electron-hole pair excitation takes place along the H-bonding direction, and a charge-transfer exciton is stabilized on the H-bonded water molecules. As a bipolar molecule, a water molecule serves both as an H-bond donor and an H-bond acceptor, and therefore both electron and hole states can be simultaneously found on a single water molecule during an excitation. As a result, large excitonic effects can be seen in Fig. {\ref{fig:figure3}} (b), which are manifested in the pronounced absorption features and large exciton binding energies of $\sim$2.2 eV in water and $\sim$3.4 eV in hexagonal ice.

\par In water, the H-bond constantly breaks and reforms on a time scale of picoseconds~\cite{Emilio2006,Kumar2007,Perakis2016}. In its liquid structure, the strength of H-bonding undergoes a significant fluctuation as well. The strength of H-bonding and its variance in water can be seen in Fig. {\ref{fig:figure3}} (c) from the broad distribution function of the proton transfer coordinate $\nu=d_{\rm OH}-d_{\rm O\cdots H}$~\cite{Ceriotti2013,Sun2018}, where $d_{\rm OH}$ denotes the length of the OH covalent bond, $d_{\rm O\cdots H}$ denotes the distance between the H atom and the O atom in the H-bond acceptor. By convention, $\nu$ has been adopted in Fig. {\ref{fig:figure3}} (a) as a descriptor to represent the H-bonding strength, where a more positive (negative) $\nu$ means a physically stronger (weaker) H-bonding strength between two neighboring water molecules. Because of its coupling to the H-bonding, it is expected that the charge-transfer exciton and its spectral signals are sensitive to its H-bonding environment in liquid. In Fig. {\ref{fig:figure3}} (c), we present the electron-hole overlap function ($\rho_{h}\times\rho_{e}$) as a function of H-bonding strength in terms of the proton transfer coordinate $\nu$ for each pair of neighboring water molecules averaged by using many configurations which are extracted from molecular dynamics simulation trajectories. The electron-hole overlap function is computed from the sum of the product of electron and hole densities obtained from BSE eigenstates whose excitation energies are located within the first absorption peak at $\sim$8 eV, and the electron-hole overlap function has been averaged over the pair of neighboring water molecules. The details of the calculations of the electron and hole densities can be found in the Section III of the Supplemental Materials. In Fig. {\ref{fig:figure3}} (c), a strong correlation between a charge-transfer exciton and its H-bonding strength can be clearly identified. When the H-bonding between two water molecules is relatively weak in Fig. {\ref{fig:figure3}} (c), the hole and electron barely overlap with each other and therefore contributes little to the observed absorption spectra. As the H-bonding become stronger, the polarization field greatly facilitates the electron-hole excitation and their overlap on the H-bonded water molecule yielding a significant increase in optical oscillator strength. Our analysis shows that the first spectral peak is dominated by charge-transfer excitons on pairs of water molecules with nearly ideal H-bonding strength comparable to the H-bonding strengths in hexagonal ice.

\subsection{Collective excitation of charge transfer excitons on water wires}
\par
\begin{figure*}[t!]
	\setlength{\abovecaptionskip}{0.cm}
	\centering
	\includegraphics[width=6.2in]{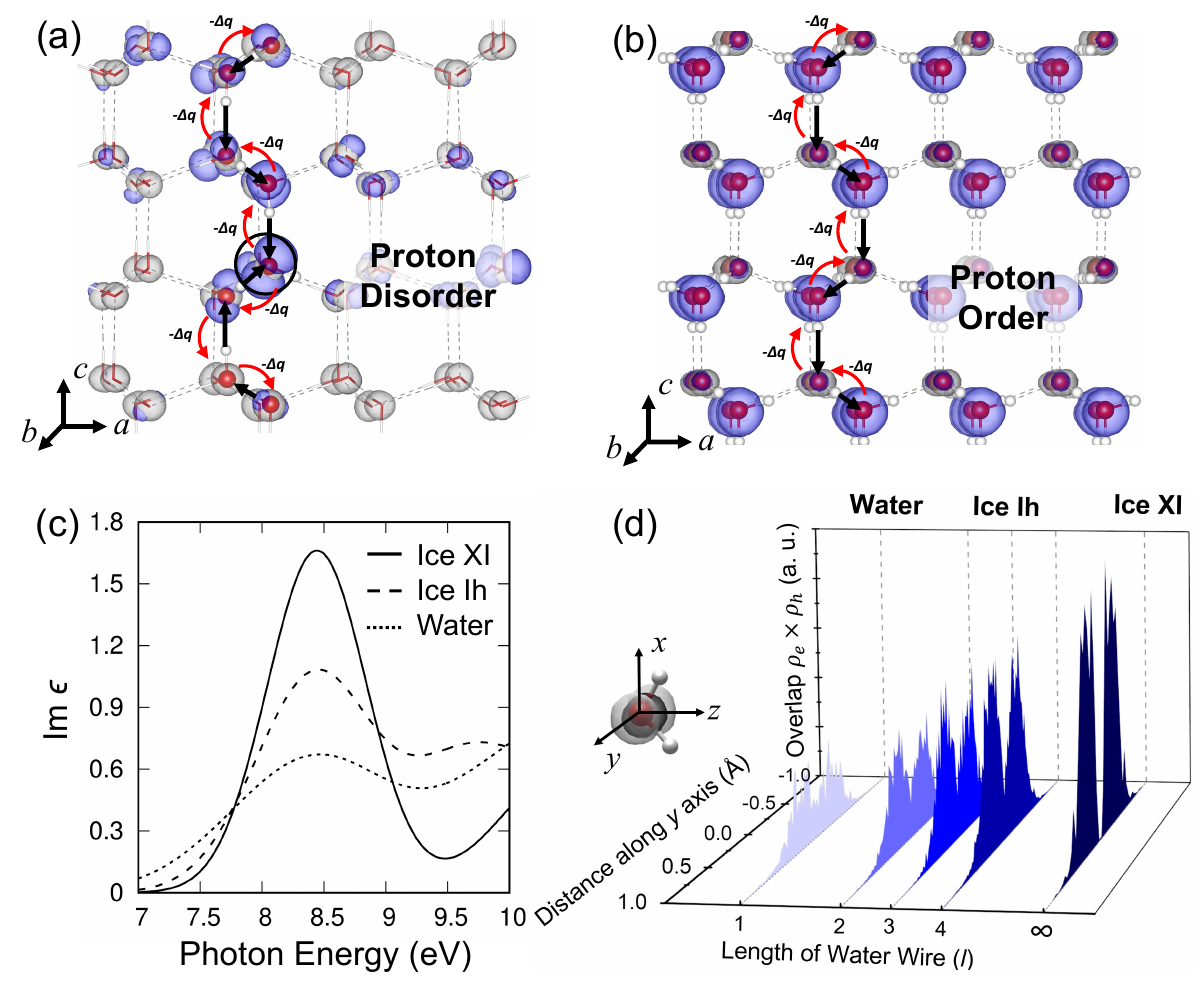}
	\caption{\label{fig:figure4} 		
		Exciton corresponding to the first peak of optical spectra and molecular structure of Ice Ih (a) and Ice XI (b). The real-space plot is the averaged electron and hole density of the exciton. The blue color denotes the hole density, while the grey color denotes the electron density. The black arrows indicate the direction of the water wire. The red arrows indicate the direction of the charge transfer. The black circle, shown in (a) marks the position where the proton disorder happens. (c) The first peak of the theoretical optical spectra of ice XI (solid line), Ice Ih (dashed line) and liquid water (dotted line). (d) Three-dimensional plot of the overlap between hole density and electron density ($\rho_{h}\times\rho_{e}$) of the excitons in the first peaks of spectra of water, Ice Ih and Ice XI, with respect to the length of the water wire. The $x$-$z$ plane is parallel to the water molecule. The zero point of the $y$ axis is set at the position of the oxygen atom. }
	\vspace{-1.0em}
\end{figure*}

  Under ambient pressure, the crystalline structure of regular ice Ih can be viewed as an assembly of corrugated oxygen bilayers stacked on top of each other as shown in Fig. {\ref{fig:figure4}} (a). In ice Ih, the oxygen atoms are located on lattice sites satisfying hexagonal symmetry, while, the protons take nearly random positions that are allowed by the \textit{ice rule}~\cite{Bernal1933,Rahman1972}, which is referred to as proton disorder in the field~\cite{Jackson1997,Kuo2005}. In ice, all neighboring water molecules are H-bonded, and the bonding strength of $\nu \simeq -0.70$ in ice is much stronger than the average bonding strength of $\nu \simeq -0.85$ in liquid water. These intact H-bonds not only construct an extended tetrahedral structure but also become interconnected among themselves and form persistent water wires on their underlying H-bond network. As shown in Fig. {\ref{fig:figure4}} (a), we present a typical water wire in regular ice. Along this water wire, five water molecules on the top three bilayers are threaded by four consecutive H-bonds in a zigzag chain. Because of the highly directional nature of H-bonding, a water wire can be defined to be a proton ordered chain if all the participating H-bonds adopt the same bonding direction pointing from donor molecule to acceptor molecule along the wire, or vice versa. As such, the above prototypical water wire can be assigned with an ordering length ($l=4$) from the number of constituent H-bonds and with an ordering direction which is along the $c$-axis of the hexagonal lattice. The ordering of the above water wire is interrupted by a proton disorder occurring at the bottom bilayer as denoted by the black circle in Fig. {\ref{fig:figure4}} (a). Therefore, only finite sized proton ordered water wires exist, whose lengths and locations are randomly distributed in ice Ih.  

\par In ice Ih, the presence of water wires greatly enhances excitonic effects at the main peak in the optical absorption spectrum. On a proton ordered water wire, the H-bonds are connected head to tail along its ordering direction, and each water molecule in ice carries an electric dipole $\sim$3.09 $D$ along a direction roughly bisecting its covalent bond angle.  In compliance with the ordering, the water molecules on a wire also position themselves in a structured configuration. As a result, the axes of the individual electric dipoles of water molecules on the wire are aligned as well, yielding a polar ordering along the $c$-axis as shown in Fig. {\ref{fig:figure4}} (b). Because of the proton disorder, the polar ordering occurs on water wires of random sizes and random direction in regular ice. Therefore, no ferroelectricity exists~\cite{Tajima1982,Xingcai1998,Lasave2021a}, and ice Ih remains paraelectric. Nevertheless, along an ordered water wire, a local polarization field arises. In ice Ih, the charge transfer exciton is not only promoted by its intact H-bonds but also further stabilized by the local polarization field along the wire. As a result, the main peak of the optical absorption spectrum in ice Ih is mainly characterized by a collective exciton excitation of charge transfer type on a  proton ordered water wire as shown in Fig. {\ref{fig:figure4}} (c). In this collective excitation, a relay of electron-hole pairs takes place. For a water molecule along the wire, a hole state is generated at the $1b_1$ valence orbital centered on oxygen atom by passing electrons onto the $4a_1$ empty orbital at the hydrogen atom of one of the neighboring molecules, and simultaneously the empty state of the same molecule receives electrons from the other neighboring molecule as shown in Fig. {\ref{fig:figure4}} (b). Here, the proton ordered water wire plays a key role in stabilizing the collective charge-transfer exciton, since its polarization field tilts the energy landscape in favor of formation of an electron and a hole on the H-bond acceptor and donor molecules, respectively, and creates a potential that reduces recombination. In the liquid water, the H-bond constantly breaks and reforms. Therefore, it cannot contain any long water wire ($l>2$) , and the number of water wires stays roughly constant with respect to temperature, since they are dominated by pairs of molecules forming H-bonds at all temperatures. Indeed, as shown in Fig. S4 of the Supplemental Material, the first peak of the optical spectra in liquid water at different temperatures is almost identical. Since, the polarization field increases with the ordering length of a water wire before reaching a limit (see Section VIII of the Supplemental Material), the collective exciton is expected to be more stabilized on a water wire with longer ordering length. Indeed, the overlap density of electron-hole pairs clearly increases when ordering length ($l$) increases as plotted in Fig. {\ref{fig:figure4}} (d), which correlates with an enhanced exciton oscillator strength. Consistently, the collective excitation on water wires in ice Ih results in a stronger binding energy of $\sim$3.4 eV compared with that in water of $\sim$2.2 eV as well as a more pronounced absorption intensity as shown in Fig. {\ref{fig:figure4}} (c).

\par At temperatures below 80 K, ice XI, a proton ordered hexagonal phase~\cite{Tajima1982,Lasave2021a}, becomes energerically more stable relative to ice Ih. In the absence of proton disorder, the water wires in ice XI develop a long-range ordering length ($l = \infty$) along the $c$ axis. On this ordered water wire with ($l = \infty$), the aligned electric dipoles maximize the polarization field and yield a ferroelectricity with $P\simeq21 {\rm \mu C/cm^2}$ along the $c$ axis of the hexagonal crystal. Not surprisingly, excitonic effects are also maximized, which is evidenced by the large increase in the magnitude of the electron-hole overlap density at $l = \infty$ in Fig. {\ref{fig:figure4}} (d). As expected, the binding energy of the main peak increases to $\sim$3.6 eV relative to $\sim$3.4 eV in ice Ih, and its intensity is also the most pronounced among all the solid and liquid phases of water under current investigation as shown in Fig.  {\ref{fig:figure4}} (c).

\section{Conclusion} 
\par Based on theoretical calculations of the optical absorption spectra of liquid water and ices within $ab$ $initio$ many-body perturbation theory, we elucidate the structural signatures of the H-bond network in experimentally observed spectral features. The main peak in the optical spectra is a charge transfer type exciton and is strongly coupled to the H-bonding strength. In water, a relatively strong H-bonding under thermal fluctuation promotes a strong excitonic effect with a pronounced oscillator strength. In ice, the presence of water wires further enhances the electron-hole interaction via a collective exciton excitation facilitated by the polarization field from the aligned electric dipoles. The collective exciton excitation is also sensitive to the ordering length of the water wire. As demonstrated in water, regular ice, and ice XI, the increasing oscillator strength of the main exciton peak can be interpreted as a signature of short range, intermediate range and long range ordered water wires. Based on this approach, the water wires, in physiological systems~\cite{Freier2011,Paulino2020,Kratochvil2023}, and under nano-confinement~\cite{Gerhard2007,Li2021,kapil2022first,ravindra2024}, and in the rich phase diagram of solid ices~\cite{Bridgman1912,Bridgman1937,Petrenko2002,Salzmann2009,Zhanglinfeng2021} can be probed and interpreted by optical absorption spectroscopy, opening pathways for the optical detection and characterization of energy, charge, and information transfer in aqueous environments.

\vspace{2em}

\section{Acknowledgments} 
\par This work was supported by the Computational Chemical Center: Chemistry in Solution and at Interfaces funded by The DoE under Award No. DE-SC0019394. D.Y.Q. was supported by the National Science Foundation (NSF) Condensed Matter and Materials Theory (CMMT) program under Career Grant Number DMR-2337987 and Grant Number DMR-2114081. Development of the BerkeleyGW code was supported by Center for Computational Study of Excited-State Phenomena in Energy Materials (C2SEPEM) at the Lawrence Berkeley National Laboratory, funded by the U.S. Department of Energy, Office of Science, Basic Energy Sciences, Materials Sciences and Engineering Division, under Contract DE-AC02-05CH11231. The computational work used resources of the National Energy Research Scientific Computing Center (NERSC), a U.S. Department of Energy Office of Science User Facility operated under Contract No. DE-AC02-05CH11231. This research used resources of the Oak Ridge Leadership Computing Facility at the Oak Ridge National Laboratory, which is supported by the Office of Science of the U.S. Department of Energy under Contract No. DE-AC05-00OR22725. This research includes calculations carried out on Temple University’s HPC resources and thus was supported in part by the National Science Foundation through major research instrumentation grant number 1625061 and by the US Army Research Laboratory under contract number W911NF-16-2-0189.

\vspace{1em}

\bibliography{optical_spectra}

\end{document}


\title{Supplemental Material for ``Optical absorption spectroscopy probes water wire and its ordering in a hydrogen-bond network''}
\author{Fujie Tang} 
\affiliation{Department of Physics, Temple University, Philadelphia, PA 19122, USA}
\affiliation{Pen-Tung Sah Institute of Micro-Nano Science and Technology, Xiamen University, Tan Kah Kee Innovation Laboratory (IKKEM), Xiamen, 361005, China}

\author{Diana Y. Qiu}
\thanks{$^\ast$Corresponding author. Email: diana.qiu@yale.edu}
\affiliation{Department of Mechanical Engineering and Materials Science, Yale University, New Haven, CT 06520, USA}

\author{Xifan Wu}
\thanks{$^\ast$Corresponding author. Email: xifanwu@temple.edu}
\affiliation{Department of Physics, Temple University, Philadelphia, PA 19122, USA}
\date{\today}

\maketitle
\section{Details of Molecular Dynamics Simulations}
\par In the main text, we reported optical spectra of liquid water and ice Ih with the GW plus Bethe-Salpeter equation (GW-BSE) method. Here we will discuss how we obtain the molecular configurations for the calculations. First, the configurations of liquid water used for the spectral calculations were adopted from our previous studies, reported in Ref.~\cite{Tang2022}. We will briefly introduce the details for it. The configurations are extracted from a path-integral deep potential molecular dynamics (PI-DPMD) simulation trajectory. The deep potential model was trained on the density functional theory (DFT) data obtained with the hybrid strongly constrained and appropriately normed (SCAN0) functional~\cite{zhang2021jpcb}. The cubic cells contain 32 water molecules while the size of supercell was adjusted to have the same density as experiment. PI-DPMD simulation was performed in the NVT ensemble at $T = 300 K$ and $T = 365 K$ using periodic boundary conditions with the cell sizes fixed at 9.708 $\rm \AA$. The Feynman paths were represented by 8-bead ring polymers coupled to a color noise generalized Langevin equation thermostat (i.e. PIGLET)~\cite{Ceriotti2009,Ceriotti2012}. The total length of PI-DPMD simulation used well equilibrated $\sim$500 ps long trajectories. After obtaining the MD trajectory, we used a score function~\cite{Tang2022} to pick out two representative snapshots for the following GW-BSE calculation.

\par For the configurations used in the ice Ih calculation, we first generated the 8 initial geometries of ice Ih following the ice rules~\cite{Bernal1933,Rahman1972} with different random seeds, in which the protons are in a disordered structure. The supercell contains 64 water molecules with a size of $a = 8.9873$ $\rm \AA$, $b = 15.5665$ $\rm \AA$, and $c = 14.6763$ $\rm \AA$. We carried out a DPMD simulation at $T = 80 K$ instead of PI-DPMD simulation to obtain the molecular configuration of ice used for the GW-BSE calculation. \FT{The reason PI-DPMD simulation was not used for ice is that, an 8-bead simulation based on the Generalized Langevin Equation (GLE) thermostat does not achieve convergence for ice at 80 K. Achieving convergence may require 128 or more beads~\cite{yang2024deuteration}, which would introduce prohibitively high computational costs for performing the GW-BSE calculations on ice. Consequently, we chose to use classical simulations to sample multiple proton-disordered configurations in ice. Nonetheless, we anticipate that the spectral amplitude is not highly sensitive to NQE, as evidenced by previous X-ray absorption spectroscopy calculations for liquid water~\cite{Sun2018} and our optical spectra for liquid water. NQE, however, may influence the bandgap of ice Ih, as recently reported~\cite{berrens2024nuclear}. For ice, as a crystalline structure, the spectral dependence on the NQEs is expected to be even smaller than liquid water.} The deep potential model was trained at the level of SCAN0 theory~\cite{zhang2021jpcb}. 8 different snapshots are extracted from trajectories for the following GW-BSE calculations of ice. For ice XI, the number of water molecules and cell size are set to the same as the setting used in the ice Ih. We optimized the structure of Ice XI at $0 K$ at the SCAN0 level of theory before the GW-BSE calculation. 

\section{Details of Ground State Calculation and GW-BSE Calculation}
We first constructed the starting point wavefunctions for our GW-BSE calculation using DFT at the level of the hybrid generalized gradient approximations (GGA) of Perdew, Burke and Ernzerhof with $25\%$ of Hartree–Fock exchange energy (PBE0)~\cite{Perdew1996}, as implemented in the Quantum ESPRESSO package~\cite{Giannozzi2017}. We chose PBE0 to reduce the overestimated charge transfer effects induced by the ground state theory~\cite{cohen2008i,cohen2012c}. \FT{Note that the hybrid XC functionals mitigate these artifacts, but they cannot completely eliminate them. Consequently, the calculated peaks still exhibit a blue-shift relative to the experimental results}. Multiple-projector norm-conserving pseudopotentials that match the potentials for oxygen and hydrogen were generated by using the ONCVPSP package~\cite{hamann2013}. \FT{All the calculations are done with the $\Gamma$-point only approximation. Note that the previous studies~\cite{Prendergast2005,Chen2016,Tang2022} have shown that the $\Gamma$-point only approximation with a 32 H$_2$O cell is sufficient to reproduce experimental optical absorption spectra, while it may be insufficient to reproduce the unoccupied states features.} The plane-wave kinetic energy cutoff was set to 85 Ry. The GW-BSE calculation is done with the BerkeleyGW package~\cite{Deslippe2012}. \FT{Note that temperature effects are accounted for only in the atomic configurations and are not included in the electronic structure calculations}. \FT{We constructed a static inverse dielectric matrix within the random phase approximation (RPA) and utilize the Hybertsen-Louie generalized plasmon-pole (HL-GPP) model to extend the dielectric response to non-zero frequencies in the GW calculation~\cite{Deslippe2012}. The same static inverse dielectric matrix is used for the BSE calculations}. The quasi-particle energies of the systems were obtained with the one-shot G$_0$W$_0$ approximation, where quasi-particle corrections were computed for 1024 empty states (liquid water) and 2048 empty states (ice Ih/XI) above the conduction band minimum. The screened cutoff is set to be 5 Ry. By extrapolating the quasiparticle gap with the number of empty states, we estimate that the quasiparticle gap at the G$_0$W$_0$ level is converged within 1 eV. Then, we apply a rigid shift of the absorption spectrum at the BSE level to compensate for the underconvergence of the quasiparticle gap. At the BSE level, we aligned the theoretical spectra respect to the experimental data by shifting the theoretical spectra by blue-shifting the theoretical spectrum by $\sim$0.6 eV (liquid water) and $\sim$0.8 eV (ice), respectively. The BSE Hamiltonian was constructed in a basis with 128 valence band states and 192 conduction band states for liquid water, and 192 valence band states and 240 conduction band states for ice, which are enough to cover the energy range in which we are interested. For all the calculations, we solved for the electron-hole excitations within the GW-BSE approach in the Tamm-Dancoff approximation (TDA)~\cite{Rohlfing2000}. All spectra are broadened with an empirical Gaussian broadening of 0.4 eV.

\section{Detailed Analysis of the Electron and Hole for Each Exciton State in Real Space}

\par To show the localization of the excitons in the real space, we define an average electron(hole) density by integrating out the position of the hole (electron)~\cite{Kshirsagar2021}. The average electron and hole densities for each exciton $\mathcal{S}$ are defined as~\cite{Kshirsagar2021}: 
\begin{align}
	\rho^{\mathcal{S}}_{e}(\bm{r}_{e})=\int\left|\Psi^{\mathcal{S}}(\bm{r}_{h},\bm{r}_{e})\right|^2d\bm{r}_{h}\\
	\rho^{\mathcal{S}}_{h}(\bm{r}_{h})=\int\left|\Psi^{\mathcal{S}}(\bm{r}_{h},\bm{r}_{e})\right|^2d\bm{r}_{e}
\end{align}
The electron/hole average densities can be interpreted as generalized conduction/valence orbitals for a given exciton. They are interesting because they allow to connecting excitons to the corresponding molecular structures by looking at the relative position of the electron/hole densities.

\par
\begin{figure*}[ht]
	\setlength{\abovecaptionskip}{0.cm}
	\centering
	\includegraphics[width=6.0in]{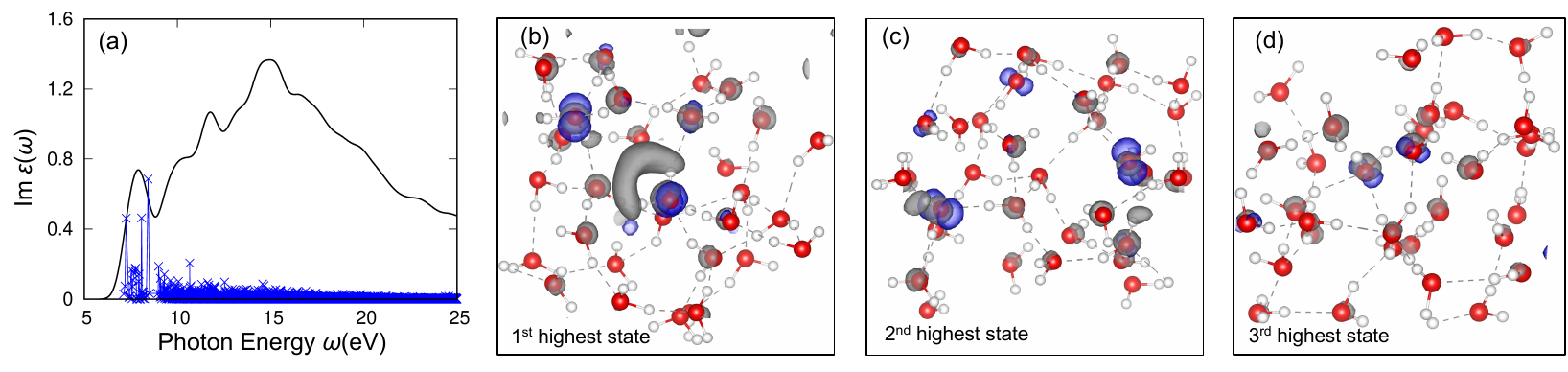}
	\caption{\label{fig:figures1} 		
		(a) Theoretical optical spectra of the liquid water (black line) and corresponding eigenstates (blue line). The calculation is done with a single snapshot. (b-d) The averaged hole and electron densities of the top three exciton states with highest absorption cross sections in the real space. The blue color denotes the averaged hole density, while the black color denotes the averaged electron density.
	}
	\vspace{1.0em}
\end{figure*}

 In the main text, we argue that that the first peak of the optical spectra for liquid water is dominated by the water molecule pair which forms a strong H-bond. Here we show one calculated spectra and corresponding exciton plots in the real space, in Fig.~\ref{fig:figures1}. As one can see, the first absorption peak is indeed dominated by the several excitons states with high absorption cross sections. 

\section{The Decomposed Spectra Based on Restricted Subspace Method}
\par In the main text, we directly show absorption spectra decomposed according to contribution from specific band-to-band transitions. The decomposed spectra are calculated by solving the BSE on a restricted subspace containing only specific bands. In this section, we will rigorously discuss the decomposition. In the TDA in the electron-hole basis, the BSE describing the electron-hole excitations can be written as an eigenvalue problem: 
\begin{equation}
	\mathcal{H}^{\bf BSE}\phi_{\mathcal{S}}=\Omega_{\mathcal{S}}\phi_{\mathcal{S}},
\end{equation}
where $\Omega_{\mathcal{S}}$ are eigen-energies corresponding to the exciton excitation energy and $\phi_{\mathcal{S}}$ are the exciton eigenvectors. We can separate the basis into subspaces containing contributions from specific orbitals. In this way, the BSE Hamiltonian can be written as:
\begin{equation}
\mathcal{H}^{\bf BSE}=\left(\begin{array}{cc}
	\mathcal{H}_{ss}&\mathcal{H}_{sp}\\
	\mathcal{H}_{ps}&\mathcal{H}_{pp}
\end{array}\right),
\end{equation}
where $s$ denotes the subspace in which the hole belongs to the valence states with $2a_1$ character, while $p$ denotes the subspace in which the hole belongs to the valence states with $p\in(1b_1,3a_1,1b_2)$ character. The hole contributions in the $p$ space, can be further separated into hole contributions with $1b_1$, $3a_1$, $1b_2$ character, resulting in three subspaces: $\mathcal{A}:\mathcal{T}_{{1b_1}\rightarrow c}$, $\mathcal{B}:\mathcal{T}_{{3a_1}\rightarrow c}$ and $\mathcal{C}:\mathcal{T}_{{1b_2}\rightarrow c}$, where the $c$ denotes the conduction band states. As such, the full Hilbert space of the BSE effective Hamiltonian, can be written as $\mathcal{H}^{\bf BSE}=\mathcal{H}_s\oplus\mathcal{H}_p$, where $\mathcal{H}_p=\mathcal{H}_{\mathcal{A}}\oplus\mathcal{H}_{\mathcal{B}}\oplus\mathcal{H}_{\mathcal{C}}$. 

\par
\begin{figure*}[ht]
	\setlength{\abovecaptionskip}{0.cm}
	\centering
	\includegraphics[width=3.0in]{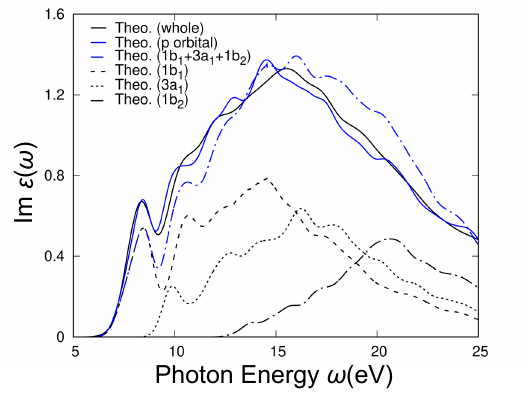}
	\caption{\label{fig:figures2} 		
		The decomposed optical spectra of liquid water based on a restricted subspace method: the whole spectrum (black), the spectrum including the $p$ orbital contributions only (blue), the sum of the spectra of decoupled $1b_1$, $3a_1$, and $1b_2$ orbital contributions (blue dash-dotted line), the spectra of $1b_1$ orbital contributions only (dash line), the spectra of $3a_1$ orbital contributions only (dotted line), and the spectra of $1b_2$ orbital contribution only (dash-dotted line). The calculation is done with a single snapshot. 
	}
		\vspace{1.0em}
\end{figure*}

 In the full calculation, all electron-hole pairs contribute to the spectra, shown as a black line in Fig.~\ref{fig:figures2}. By decomposing the contributions by solving the BSE on a subspace, we find that the full spectrum is well approximated by the solution on a subspace containing only the $p$ orbital electrons near the Fermi level (i.e. the $1b_1$, $3a_1$, $1b_2$ bands). There are no contributions from inner valences $2a_1$ bands, as we discussed in the main text. This is because that transition takes place at a higher energy. From this analysis, we see that the coupling of the $s$ subspace and $p$ subspace is negligible, due to the fact that $s$ and $p$ orbitals are orthogonal to each other. This can be seen in Fig.~\ref{fig:figures2}, where the full spectrum is well-approximated by the spectrum with $p$ orbital contribution only. All three subspaces of the $p$-orbital ($1b_1$, $3a_1$, and $1b_2$) contribute significantly to the optical spectra as shown in the Fig.~\ref{fig:figures2}. However, the coupling between these subspaces is relatively unimportant, as the spectral lineshape remains qualitatively similar to the full spectrum if we simply sum the absorption spectra on each subspace; that is, we can approximate $\mathcal{H}^{\bf BSE}=\mathcal{H}_{\mathcal{AA}}+\mathcal{H}_{\mathcal{BB}}+\mathcal{H}_{\mathcal{CC}}$, where the contributing electron-hole transitions are $\mathcal{T}=\mathcal{T}_{{1b_1}\rightarrow c}+\mathcal{T}_{{3a_1}\rightarrow c}+\mathcal{T}_{{1b_2}\rightarrow c}$. \FT{Note that in Fig. 2(c) of the main text, the 'Overall' spectrum is lower in magnitude than the $\mathcal{T}_{{1b_2}\rightarrow c}$ contribution around 22.5 eV due to destructive interference between different subspaces.}

\section{Determination of the Exciton Binding Energy}

\par
\begin{figure*}[ht]
	\setlength{\abovecaptionskip}{0.cm}
	\centering
	\includegraphics[width=3.0in]{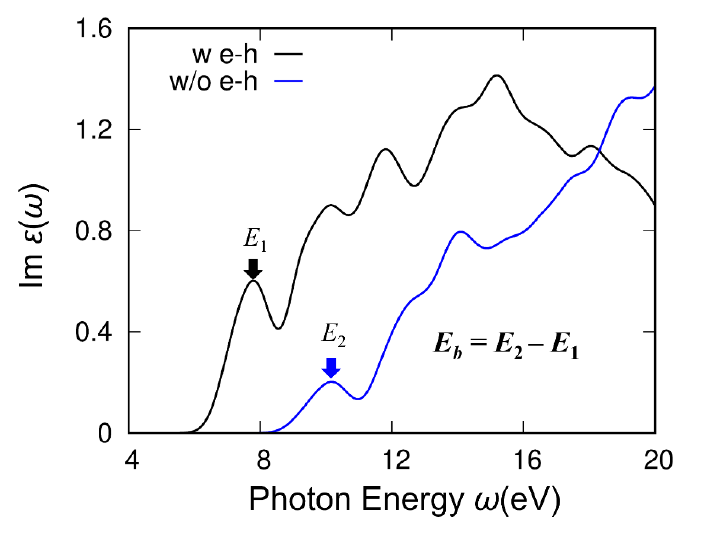}
	\caption{\label{fig:figures3} 		
		Theoretical optical spectra of the liquid water calculated with GW-BSE (including e-h interaction (w e-h)) and GW-RPA (without e-h interaction (w/o e-h)) approach. The calculation is done with a single snapshot.
	}
			\vspace{1.0em}
\end{figure*}

 To determine the exciton binding energy, we plot the absorption spectrum obtained using the BSE and the corresponding spectrum obtained in the GW-RPA approach, see Fig.~\ref{fig:figures3}. We then determine the exciton binding energy Eb from the separation between the first peak in the BSE and RPA spectra, following the same definition used in Ref.~\cite{Nguyen2019}. After obtaining the exciton binding energy for each individual snapshot, we average the values to get the averaged binding energy $E_b$ for liquid water: $E_b =$ $\sim2.2$ eV, ice Ih: $E_b =$ $\sim3.4$ eV and XI: $E_b =$ $\sim3.6$ eV. \FT{Here, we calculated the exciton binding energy from the eigenvalues of the BSE Hamiltonian as well as the spectral onsets: liquid water: $E_b =$ $\sim2.4$ eV, ice Ih: $E_b =$ $\sim3.1$ eV and ice XI: $E_b =$ $\sim3.6$ eV. The corresponding bandgap of liquid water is $E_g =$ $\sim8.8$ eV, ice Ih: $E_g =$ $\sim9.9$ eV and ice XI: $E_g =$ $\sim10.0$ eV. In this case, since we are looking at disordered structures, once can expect some broadening in both the distribution of GW energies and exciton energies due to the disorder, and the exact energies may change depending on the snapshot used. Hence, simply looking at the difference between the GW bandgap and exciton eigen-energies, which is accurate for crystalline solids, may be less physically meaningful than comparing averaged spectral onsets. Nevertheless, the conclusion remains unchanged with different definitions of the binding energy.}

\section{Temperature Dependence of the Optical Spectra of Liquid Water}
\par
\begin{figure*}[ht]
	\setlength{\abovecaptionskip}{0.cm}
	\centering
	\includegraphics[width=3.0in]{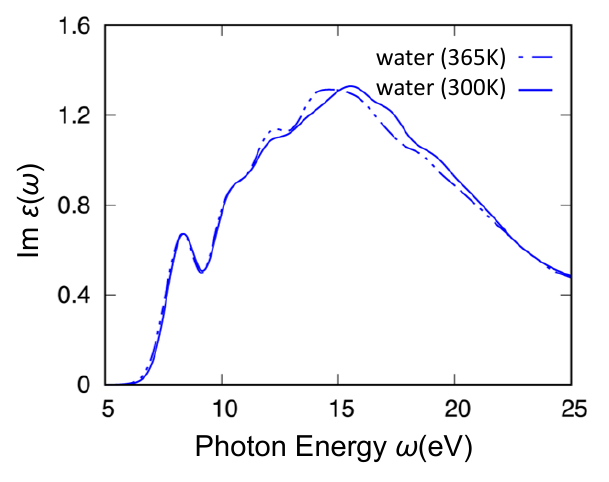}
	\caption{\label{fig:figures4} 		
		Theoretical optical spectra of liquid water calculated with PI-DPMD simulation at 300 K (solid line) and 365 K (broken line). The calculations are done with two snapshots (total of 16 structures for each temperature).
	}
			\vspace{1.0em}
\end{figure*}

 In the main text, we have argued that the first peak of the optical spectra of liquid water is the signature of the short-range H-bond network. The height of the first peak is sensitive to the length of the water wires. In liquid water, the H-bond constantly breaks and reforms on a time scale of picoseconds~\cite{Emilio2006,Kumar2007,Perakis2016}. Therefore, the length of the water wires in liquid water at different temperatures should be $l = 1$ at any temperature, as only two water molecules consistently form H-bond pairs. Indeed, the first peaks of the optical spectra of liquid water at different temperatures are identical, as shown in Fig.~\ref{fig:figures4}. It further demonstrates our argument that the first peak of the optical spectra is the signature of the water wire.

\FT{\section{Absorption Coefficient Spectra of the Liquid Water and Ice}
\par In the main text, we presented the imaginary part $\varepsilon_2$ of the dielectric function of the liquid water and ice. Here, we presented the corresponding real part $\varepsilon_1$ of the dielectric function of the liquid water and ice, shown in Fig.~\ref{fig:figures_realpart}. As one can see, our calculations show good agreement with the experimental spectra as well.

\begin{figure*}[ht]
	\setlength{\abovecaptionskip}{0.cm}
	\centering
	\includegraphics[width=3.0in]{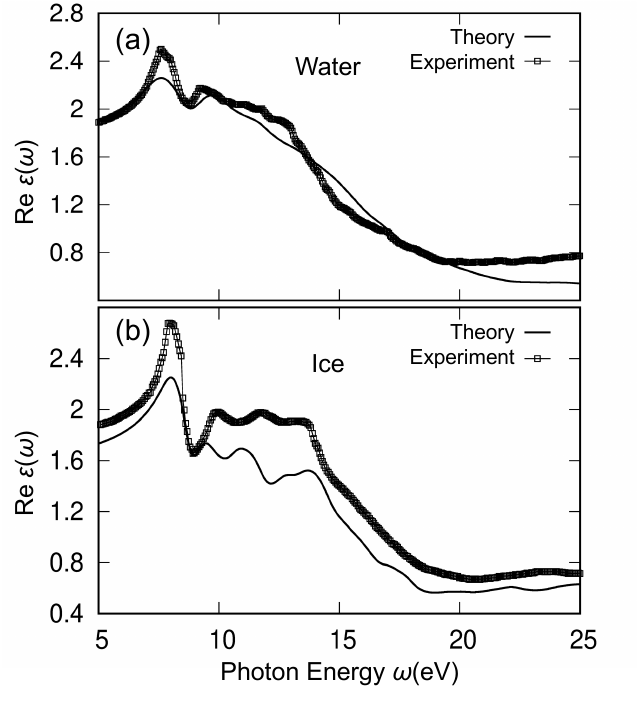}
	\caption{\label{fig:figures_realpart} 		
		\FT{The experimental and theoretical spectra of the real part of the dielectric function of (a) liquid water (a) and (b) ice. The theoretical optical spectrum of liquid water was generated by using the atomic configuration (32 H$_2$O) from a PI-DPMD simulation at 300 K, while the ice spectrum was generated by using the atomic configuration (64 H$_2$O) from a DPMD simulation for ice Ih at 80 K. The theoretical spectra are aligned with the shift as the same as the ones used in the imaginary spectra with a blue shift of $\sim$0.6 eV (liquid water) and $\sim$0.8 eV (ice), respectively. The experimental data are taken from Refs.~\cite{Hayashi2015} and ~\cite{Koichi1983} for water and ice, respectively.}
	}
	\vspace{1.0em}
\end{figure*}

Using the knowledge of the complex dielectric constant, one can calculate the absorption coefficient $\alpha(\omega)$ using the following equation:
\begin{equation}
\alpha(\omega)=\frac{\omega}{c}\sqrt{2(\sqrt{\varepsilon_1^2(\omega)+\varepsilon_2^2(\omega)}-\varepsilon_1^2(\omega))}.
\end{equation}
Here, we present the calculated absorption coefficient spectra $\alpha(\omega)$  of the liquid water and ice in the range from $\sim$50 nm to $\sim$170 nm, shown in Fig.~\ref{fig:figures_ac}, which is corresponding with the range from 7 eV to 25 eV. The absorption peak at around $\sim$8 eV in the imaginary spectra is corresponding to the $\sim$155 nm peak in both water and ice absorption coefficient spectra. Our calculations for all the imaginary and real parts dielectric function as well as the absorption coefficient spectra successfully reproduce the experimental observables, especially, near the range of the absorption edge.

\begin{figure*}[ht]
	\setlength{\abovecaptionskip}{0.cm}
	\centering
	\includegraphics[width=3.0in]{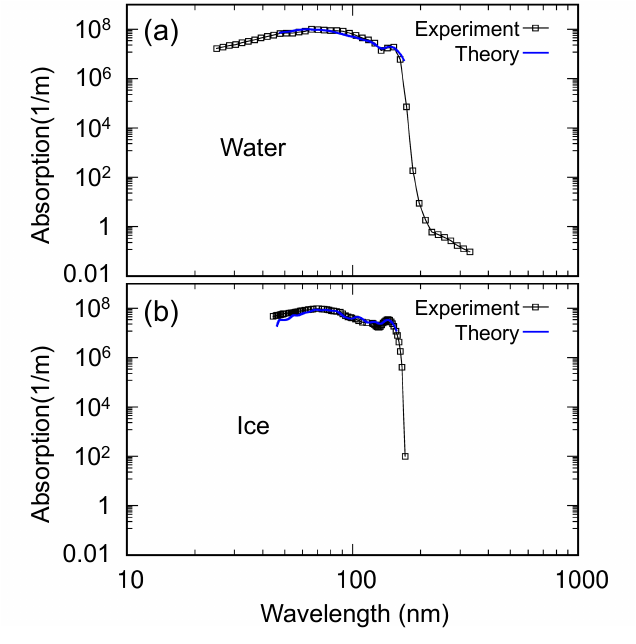}
	\caption{\label{fig:figures_ac} 		
		\FT{The experimental and theoretical absorption coefficient of (a) liquid water (a) and (b) ice as a function of the optical frequency. The theoretical spectra are in the range from $\sim$50 nm to $\sim$170 nm. The experimental data of ice is taken from Ref.~\cite{Koichi1983} and the water data is taken from Ref.~\cite{segelstein81}. }
	}
	\vspace{1.0em}
\end{figure*}
}

\section{Local Polar Fields along the Water Wire in Ice Ih}

\par In the main text, we show that the electron-hole overlap will be enhanced when the length of the water wire is longer. We argue that the effects are correlated with the local polar field along the water wire. Here, we calculate the average dipole moment of the water molecule in liquid water, ice Ih and ice XI, as well as the ones belonging to the water wire in ice Ih, shown in Tab.~\ref{tab:tables1}.  

\par Let us consider an approximation, if the excited electron (hole) density on a water molecule is proportional (inversely proportional) to the local polar field $P$, then the electron-hole overlap ($\Omega$) can be written as $\Omega=kP^2$, where the $k$ is a constant. Then, we to first order in $\Delta P$, we can approximate the relation between the change of the electron-hole overlap ($\Delta\Omega$), and the change of the local polar field ($\Delta P$) as $\Delta\Omega=2kP\Delta P$. With this first order approximation, the change in the electron-hole overlap $\Delta\Omega$ and the change of the local polar field $\Delta P$ will be linear. The above relation is solely based on the assumption that the local polar field is increases when the length of the water wire is longer. 

\par
\begin{figure*}[ht]
	\setlength{\abovecaptionskip}{0.cm}
	\centering
	\includegraphics[width=3.0in]{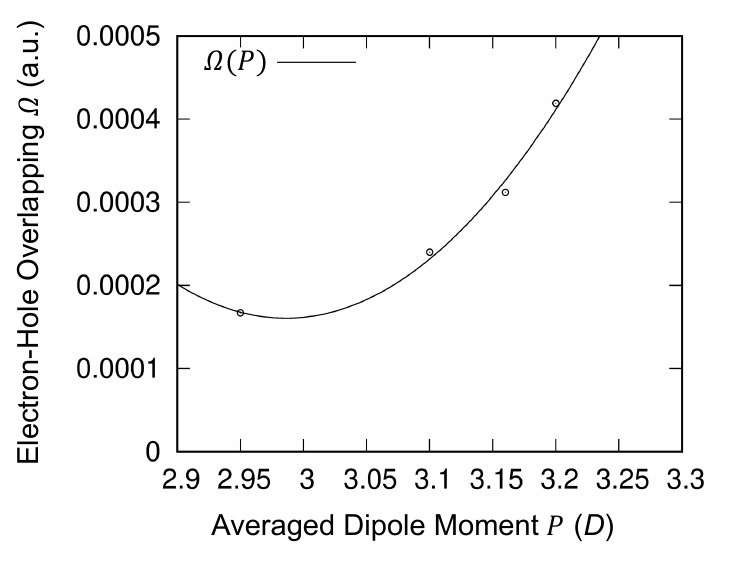}
	\caption{\label{fig:figures5} 		
		The relation between the average dipole moment $P$ and the estimated electron-hole overlapping $\Omega$. The fitting function is $\Omega(P)=0.005488*P^2-0.0327788*P+0.0491024$. Note that the overlap $\Omega$ is estimated by using the peak area of the data in Fig. 4(c) in the main text. 
	}
			\vspace{1.0em}
\end{figure*}

 To this end, we plot the relation between the electron-hole overlap observed in our calculation for the liquid water and ice Ih as a function of the local polar field, or the averaged dipole moment of the water molecule, as shown in Fig.~\ref{fig:figures5}. As one can see, indeed the relation $\Omega=kP^2$ holds for the local polar field ($P$) and the electron-hole overlap ($\Omega$).

\begin{table*}[htbp]
	\centering
	\caption{The calculated averaged dipole moment of the water molecule in liquid water, ice Ih and ice XI, as well as the ones belonging  to the water wire in the ice Ih. The calculations are done with the Wannier center methods implemented in the Quantum Espresso package~\cite{Giannozzi2017}.}
	
		\vspace{3.0em}
	\begin{tabular}{cccccccc}
		 \hline
		&       &       & liquid water & \multicolumn{3}{c}{ice Ih} & ice XI \\
		 \hline
		\multicolumn{3}{c}{\multirow{3}[1]{*}{averaged dipole moment ($D$)}} & \multirow{3}[1]{*}{2.95} & \multicolumn{3}{c}{3.09} & \multirow{3}[1]{*}{3.36} \\
		\multicolumn{3}{c}{}  &       & $l = 2$ & $l = 3$ & $l = 4$ &  \\
		\multicolumn{3}{c}{}  &       & 3.1   & 3.16  & 3.2   &  \\
		\hline
	\end{tabular}%
	\label{tab:tables1}%
\end{table*}%

\newpage

\section{Supplemental Figures}

\begin{figure*}[ht]
	\setlength{\abovecaptionskip}{0.cm}
	\centering
	\includegraphics[width=3.0in]{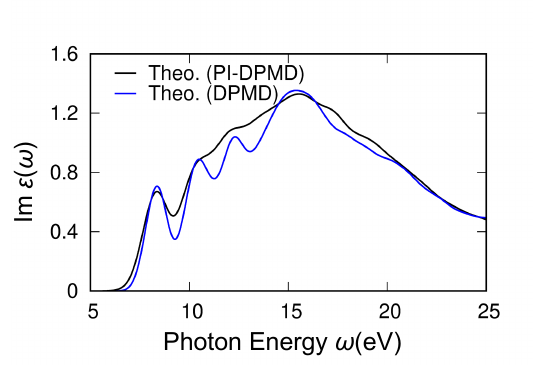}
	\caption{\label{fig:figures6} 		
		Theoretical optical spectra of liquid water calculated based on the snapshots generated by using PI-DPMD (black) and DPMD (blue) simulation. 
	}
	\vspace{1.0em}
\end{figure*}

\begin{figure*}[ht]
	\setlength{\abovecaptionskip}{0.cm}
	\centering
	\includegraphics[width=3.0in]{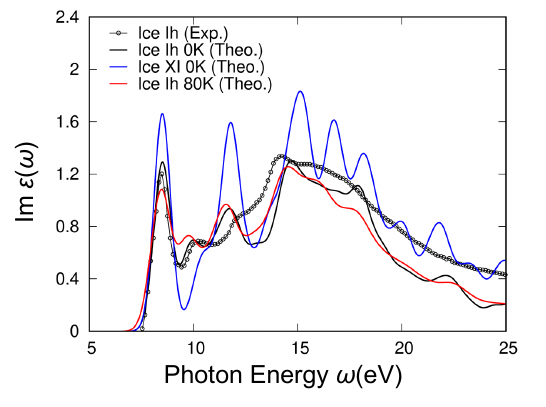}
	\caption{\label{fig:figures7} 		
		Theoretical optical spectra of ice Ih at 0 K (black line), ice Ih at 80 K (red line), and ice XI at 0 K (blue line). The experimental spectra of ice Ih (circle) is shown for comparison.  
	}
	\vspace{1.0em}
\end{figure*}

\begin{figure*}[ht]
	\setlength{\abovecaptionskip}{0.cm}
	\centering
	\includegraphics[width=3.5in]{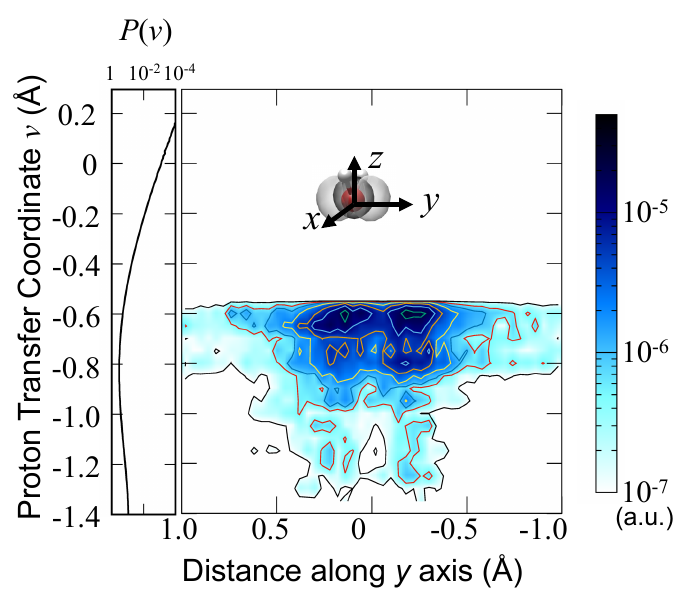}
	\caption{\label{fig:fig3c} 		
\FT{The two-dimensional contour plot of the overlap between hole density and electron density ($\rho_{h}\times\rho_{e}$) with respect to the distance along the y axis (perpendicular to the plane of the water molecule) and the proton transfer coordinate. The proton transfer coordinate distribution in the logarithmic-scale is shown for reference. }
	}
	\vspace{1.0em}
\end{figure*}

\FloatBarrier

\bibliography{SI_optical_spectra}